\documentclass[preprint2,12pt]{emulateapj}

\usepackage{amsmath}
\usepackage{natbib}
\slugcomment{Accepted for Publication in the Astrophysical Journal Letters}

\shorttitle{Ruling out X-rays from IC/CMB in 3C 273}
\shortauthors{Meyer \& Georganopoulos}

\begin{document}

\title{\emph{Fermi} rules out the IC/CMB model for the Large-Scale Jet X-ray emission of 3C 273}

\author{Eileen T. Meyer\altaffilmark{1,2}, Markos
  Georganopoulos\altaffilmark{2,3}} \altaffiltext{1}{Space Telescope
  Science Institute, 3700 San Martin Drive, Baltimore, MD 21218}
\altaffiltext{2}{Department of Physics, 
  University of Maryland Baltimore County, 1000 Hilltop Circle,
  Baltimore, MD 21250, USA} \altaffiltext{3}{NASA Goddard Space Flight
  Center, Code 660, Greenbelt, MD 20771, USA}
\begin{abstract}

The X-ray emission mechanism in large-scale jets of powerful radio
quasars has been a source of debate in recent years, with two
competing interpretations: either the X-rays are of synchrotron
origin, arising from a different electron energy distribution than
that producing the radio to optical synchrotron component, or they are
due to inverse Compton scattering of cosmic microwave background
photons (IC/CMB) by relativistic electrons in a powerful relativistic
jet with bulk Lorentz factor $\Gamma\sim 10-20$.  These two models
imply radically different conditions in the large-scale jet in terms
of jet speed, kinetic power, and maximum energy of the particle
acceleration mechanism, with important implications for the impact of
the jet on the large-scale environment. A large part of the X-ray
origin debate has centered on the well-studied source 3C 273.  Here we
present new observations from \emph{Fermi} which put an upper limit on
the gamma-ray flux from the large-scale jet of 3C 273 that violates at
a confidence greater that $99.9\%$ the flux expected from the IC/CMB
X-ray model found by extrapolation of the UV to X-ray spectrum of knot
A, thus ruling out the IC/CMB interpretation entirely for this source
when combined with previous work. Further, this upper limit from
\emph{Fermi} puts a limit on the Doppler beaming factor of at least
$\delta <$9, assuming equipartition fields, and possibly as low as
$\delta <$5, assuming no major deceleration of the jet from knots A
through D1.

\end{abstract}


\section{Introduction}
\label{sec:intro}

Large-scale jets of kpc-Mpc size have been observed in radio images of
radio-loud AGN almost since their discovery, but only more recently
has high-resolution imaging with the Hubble Space Telescope (HST) and
the \emph{Chandra} X-ray observatory shown that the knots in many of
these large-scale jets often produce significant high-energy
radiation. Since the first (serendipitous) \emph{Chandra} detection of
a large-scale X-ray jet in PKS 0637-752 \citep{chartas2000_pks0637},
several dozen have been discovered \citep[see][for a
  review]{harris06}, spanning a range from typically lower radio power
Fanaroff and Riley \citep[FR,][]{fan74} type I to more powerful FR II
type radio galaxies.

With high-resolution multi-band imaging, we are now able to build
reliable spectral energy distributions (SEDs) for the large-scale jet
(LSJ) emission, separate from the blazar core. In many cases, the
spectra of the knots appears consistent with a single synchrotron
origin from radio to X-rays, as seen in M87
\citep{wilson2002_m87,perlman05}, B2 0331+39 \citep{worrall2001_FRIs},
and 3C 31 \citep{hardcastle2002_3c31}, and several others, all notably
FR I sources.  However, in several of the more powerful (typically FR
II) sources, the X-ray spectrum in the knots is clearly much harder
and/or higher than would be consistent with the radio-optical
synchrotron spectrum, as first observed by \cite{schwartz2000_pks0637}
and \cite{chartas2000_pks0637} for PKS 0637-752.

\begin{figure*}[t]
  \begin{centering}
  \includegraphics[width=5.5in]{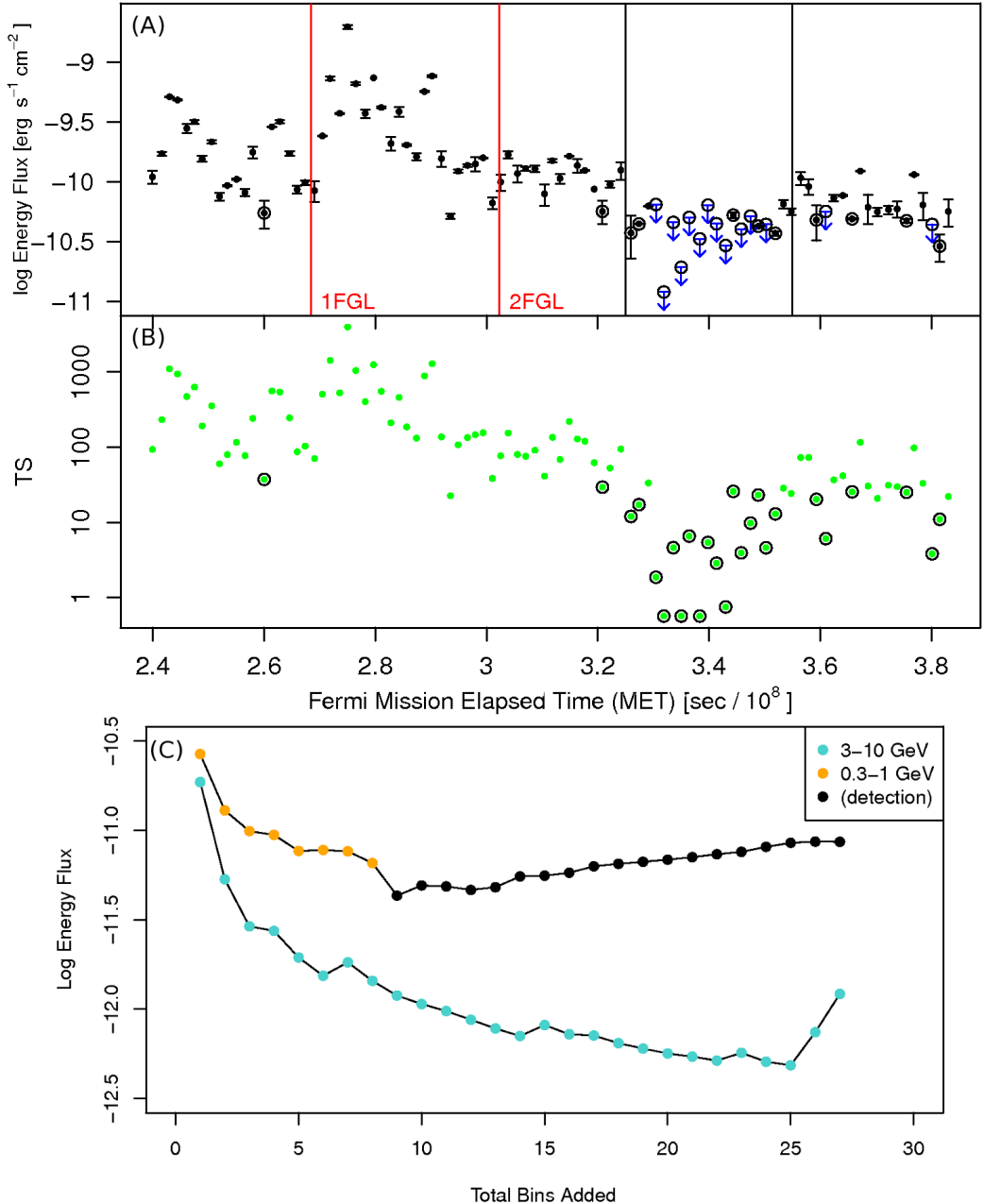}
  \caption{\emph{(A):} Total lightcurve of 3C 273 (4 August 2008 to 11
    March 2013) in bins of equal GTI time (7.5 days), showing total
    \emph{Fermi} band (100 MeV to 100 GeV) energy flux versus MET. The
    1FGL and 2FGL catalog end times are noted with red lines.
    Detections are shown as points with error bars, while upper limits
    (when the TS of the source was $<$10) are shown as arrows.  The
    circled points in both panels are the 25 lowest bins which were
    combined to give the lowest limit on the jet emission. \emph{(B):}
    The TS of each bin versus MET. \emph{(C):} The flux
    limit of the 0.3-1 GeV and 3-10 GeV bands with
    increasing integration time. Colored points indicate upper limits;
    black are detections. }
  \end{centering}

  \label{fig:lc}
\end{figure*}

Based on that finding, \cite{tavecchio2000_iccmb} and
\cite{celotti2001_iccmb} suggested that the X-rays could be due to
IC/CMB photons by relativistic electrons in the
jet\footnote{Synchrotron self-Compton has been shown to be an
  inadequate mechanism to produce the observed X-ray flux in these
  sources, unless the magnetic field in the jet is orders of magnitude
  below the equipartition value
  \citep[\emph{e.g.},][]{chartas2000_pks0637}.}.  The IC/CMB model has
since been applied to other jets with X-rays inconsistent with their
radio-optical synchrotron spectra, including the well-studied source
3C 273 \citep{sambruna2001_3c273}, and many more FR II X-ray jets
subsequently discovered (\emph{e.g.},
\citealt{sambruna2004_friis,worrall2009,mehta09}; see also the
`two-zone' IC/CMB model for PKS 1127-145 of
\citealt{aneta_pks1127}). Generally, the IC/CMB model requires that
the jet remain highly relativistic out to the location of the X-ray
knots (bulk Lorentz factor $\Gamma \sim 10-20$), point close to our
line of sight, and have an electron energy distribution (EED)
extending down to energies $\sim 1-10$ MeV, significantly lower than
the $\sim 1-10$ GeV electron energies traced by GHz synchrotron radio
emission. To produce the observed X-ray flux, IC/CMB requires high,
sometimes super-Eddington jet kinetic power
\citep{dermer04,uchiyama06}, due to the low radiative efficiency of
these electrons. Also, the small angle to the line of sight in several
cases requires Mpc-scale de-projected jet lengths, as long as the
longest radio galaxies observed \citep{dermer04,sambruna2008_xjet}.

Deep HST imaging photometry of the knots in PKS 1136-135 also reveals
`improbability' issues with the IC/CMB model, with the observed
optical polarization exceeding 30\%; applying the IC/CMB model
requires a significantly super-Eddington jet longer than a Mpc,
forming a $\sim 2.5^{\circ}$ angle to the line of sight and having a
Doppler beaming factor $\delta > $20 \citep{cara_pks1136}.

An alternative explanation for the X-rays in powerful sources is
synchrotron emission from an additional electron energy distribution
(EED) \citep[\emph{e.g.}][]{hardcastle06,
  jester06,uchiyama06}. Because the synchrotron emission mechanism is
far more efficient than IC/CMB, it does not require the high Lorentz
factors, extreme jet lengths or near-Eddington jet powers, as the
IC/CMB model does in several cases \citep{jorstad04,uchiyama06}.
However, it is not clear what physical mechanism might produce this
second EED, and in some cases the observed SED requires the
high-energy particle population to have a difficult-to-explain
low-energy cutoff at $\sim$ TeV energies, where fast cooling is
unavoidable \citep{mehta09}.

One of the best-studied LSJs is seen in the powerful nearby (z=0.158)
quasar 3C 273. Imaging in all bands reveals similar features, with a
knotty jet beginning about 12\arcsec\, from the blazar core and
extending a further 12\arcsec\, downstream. Extensive observations
with HST, \emph{Spitzer}, and \emph{Chandra} have revealed that the
knots are characterized by two spectral components, one with a cutoff
above 5$\times$10$^{13}$ Hz and a high-energy one connecting the
optical-UV and X-ray data
\citep{jester05,jester06,uchiyama06}. \cite{georganopoulos06}
(hereafter G06), showed that while the radio to X-ray SED of this
source alone cannot discriminate between the IC/CMB and synchrotron
models, \emph{gamma-ray} observations, specifically with \emph{Fermi},
may be able to do so. As discussed in G06, if the X-rays from the 3C
273 jet are due to IC/CMB, a hard, steady spectrum is also expected in
the gamma-rays by extension \citep[see also][]{sambruna04}. If \emph{Fermi} detects this emission (or
puts limits on it) at a level significantly below what is expected by
extrapolation from the X-rays, the IC/CMB model for the X-rays will be
ruled out.

The competing IC/CMB and synchrotron models imply radically different
views of the LSJ power, bulk Lorentz factor, and the efficiency of
particle acceleration, resulting in very different impacts on the host
galaxy and surrounding environment. The persistently open question of
the nature of the X-rays is critical not only for understanding jet
physics but also for our understanding of AGN activity as a feedback
mechanism in galaxy formation, yet until now no conclusive evidence
has arisen to eliminate either model.

In this paper, we analyze the gamma-rays of 3C 273 for evidence of the
expected hard, flat spectrum from IC/CMB which has been suggested as
the source of the X-rays in this and other powerful LSJs. In
Section~\ref{sec:methods}, we discuss the method of the \emph{Fermi}
data analysis and our finding that no IC/CMB emission has been
detected. In Section~\ref{sec:results} we discuss the resulting upper
limit on the IC/CMB emission along with constraints on the Doppler
beaming factor. In Section~\ref{sec:bulkgamma} we derive a limit for
the bulk Lorentz factor based on our \emph{Fermi} result.

\section{\emph{Fermi} Analysis of 3C 273}
\label{sec:methods}

We first computed the lightcurve of 3C 273 using bins of equal Good
Time Interval (GTI) time, totaling 648000 seconds (7.5 days) per bin,
corresponding to a range of 15-23 days in real time. Using the
standard pipeline tools (version v9r27p1) and the latest instrument
response function (P7SOURCE\_V6),  the flux of 3C 273 was
  calculated using (unbinned) maximum likelihood with the {\tt gtlike}
  tool.  We used a region of interest (ROI) of 7 degrees; all sources
  (29) listed in the two-year catalog \citep[2FGL;][]{nolan2012_2fgl}
  within 15 degrees of the position of 3C 273 were included in the
  initial model. In some bins, known sources which were undetected
were removed in order to gain convergence. 3C 273 was modeled as a
simple powerlaw with spectral index and normalization
free. When the test statistic (TS, roughly equivalent to
  significance squared) for 3C 273 was $<$10, we used the Fermi {\tt
    UpperLimit} tool which uses the profile likelihood method
  \cite[e.g.][]{profilestat}, freezing the other source
  parameters to generate upper limits. 

The lightcurve of 3C 273 from 4 August 2008 to 11 March 2013
(\emph{Fermi} Mission Elapsed Time (MET) 239557417 to 384684952
seconds) is shown in Figure~\ref{fig:lc}, with total flux versus the
central MET of the corresponding bin in the upper panel, and test
statistic versus the latter in the lower panel.

Previous calculations (G06) have shown that it may be possible to
detect the hard, steady component from IC/CMB by the LSJ when the
competing blazar emission is at a minimum. However, the analysis is
complicated by the fact that \emph{Fermi} lacks the spatial resolution
to resolve the LSJ separately from the blazar core, as the
\emph{Fermi} angular resolution ranges from 3.5$^\circ$ at 100 MeV
down to $\sim 0.15^\circ$ above 10 GeV, at which point it is still an
order of magnitude larger than the scale of the LSJ. As can be seen
from Figure~\ref{fig:lc}, the core appears to dominate the emission,
with significant short-term variability with timescales on the order
of the bin widths.

In order to gain the increased sensitivity of a longer integration
time on the source while avoiding times where the blazar may come `up'
during an otherwise quiescent period, we used a progressive binning
approach, in which the lightcurve bins were ordered by total
flux. Beginning with the bin with lowest flux, we then added the
next-highest bin (not necessarily contiguous) in succession and re-ran
the likelihood analysis for the combined timeframe at each
addition. The SED was divided into the five `standard' energy ranges
used in the 2FGL: 100-300 MeV, 300 MeV-1 GeV, 1-3 GeV, 3-10 GeV, and
10-100 GeV. As above, when a given energy bin found TS$<$10, an upper
limit was calculated. Overall, the flux calculations behaved as
expected: initially all bands were upper limits, which became
progressively lower as more time bins were used, up to the point where
the blazar was detected, when the flux values began increasing in the
lower-energy bins (see Figure~\ref{fig:lc}c).

The two highest energy bins (3-10 GeV and 10-100 GeV) gave only upper
limit fluxes during the entire analysis, reaching a minimum after the
25 lowest bins were analyzed together. The inclusion of bins after the
25th lowest only increased fluxes (or upper limits) in all energy
bands.  Therefore we report the 95\%, 99\% and 99.9\% upper limits on
the fluxes in the 3-10 and 10-100 GeV bands in Table~\ref{table:fermi}
using these 25 bins, in addition to the detected total fluxes in the
first three bins. We note that the background at energies $>$ 3 GeV
from nearby sources is very low as the nearest source (PKS 1217+02) is
at a distance of almost 5 times the 95\% containment radius at this energy
($\sim$0.5$^\circ$).

\begin{deluxetable}{lll}
\tablecolumns{3}
\tablewidth{0pc}
\tablecaption{\emph{Fermi} Analysis Results}
\tablehead{
\colhead{Energy Bin} & \colhead{limit} & \colhead{Energy Flux}         \\
\colhead{}           & & \colhead{erg s$^{-1}$ cm$^{-2}$} 
}
\startdata
100 - 300  MeV    &    & 1.30$\pm$0.24$\times$10$^{-11}$ \\
300 - 1000 MeV    &    & 8.50$\pm$0.78$\times$10$^{-12}$ \\
1   - 3    GeV    &    & 2.43$\pm$0.62$\times$10$^{-12}$ \\
\hline
\rule{0pt}{3ex}3   - 10   GeV  & 95\%      & $<$4.9$\times$10$^{-13}$ \\
                & 99\%      & $<$9.4$\times$10$^{-13}$ \\
                & 99.9\%    & $<$1.6$\times$10$^{-12}$ \\
\hline
\rule{0pt}{3ex}10  - 100  GeV  & 95\%      & $<$2.5$\times$10$^{-12}$ \\
                & 99\%      & $<$3.6$\times$10$^{-12}$ \\
                & 99.9\%    & $<$4.9$\times$10$^{-12}$ \\

\enddata
\label{table:fermi}
\end{deluxetable}

Alternative methods of ordering the bins are possible (such as
strictly on upper limit flux value, or by TS); these methods give
practically identical results (nearly the same ordering and a minimum
flux in the final two bins within a few percent of the above values).
The final 5-band SED points are shown in Figures 2 and 3.  It is clear
that the first three bins are a representation of the low-level blazar
SED, which is apparently peaking before the \emph{Fermi} band and
rapidly falling off in the high-energy range. The two upper limits
shown are thus upper limits for both the blazar emission and the
expected hard, steady component from IC/CMB, with the 3-10 GeV limit
being the most constraining.


\section{IC/CMB for knot A is ruled out}
\label{sec:results}

\begin{figure}
  \includegraphics[width=3.3in]{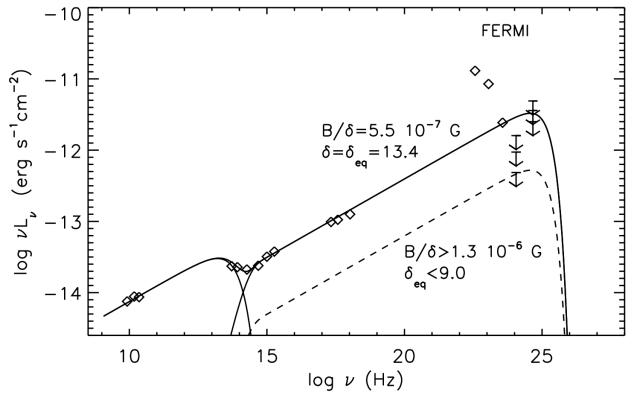}
  \caption{The SED of knot A (data from \citealt{uchiyama06} and
    \citealt{jester05,jester06}), along with the \emph{Fermi} $95\%,
    99\%$, and $99.9\%$ upper limits described in \S \ref{sec:methods}
    and Table \ref{table:fermi}.  The numerical SED calculated at
    equipartition (solid line) overproduces the 3-10 GeV $99.9\%$
    \emph{Fermi} upper limit, ruling out the IC/CMB model for the
    X-ray emission of knot A.  The broken line is the highest level
    the IC/CMB component can have without violating the $95\%$ 3-10
    GeV band \emph{Fermi} upper limit.}
  \vspace{15pt}
  \label{knotA}
\end{figure}

In the IC/CMB model, the GeV
emission is predetermined by the requirement that IC/CMB emission
gives the observed X-ray flux. 
Consider the synchrotron SED of knot A (Figure \ref{knotA}).
The synchrotron emitting electrons will unavoidably produce an IC/CMB component  (G06) identical to the synchrotron
one but with a shift in peak frequency
\begin{equation}
{\nu_c \over \nu_s}={ 2\pi m_e c(1+z) \nu_{0}\over e (B /\delta) }
=6.6 \times 10^4  {\delta \over B/B_0}=6.6 \times 10^8 \; \delta^2,
\label{eq:f}
\end{equation}
and a shift in peak luminosity
\begin{equation}
{L_c \over L_s }={32 \pi U_{0} (1+z)^4  \over 3 (B/\delta)^2}=2.5 \times 10^{-11}  \left({\delta \over B/B_0}\right)^2 =2.5 \times 10^{-3} \;\delta^4,\label{eq:L}
\end{equation}
where $\nu_c$ and $\nu_s$ are the peak EC and synchrotron frequencies,
$L_c$ and $L_s$ are the peak EC and synchrotron luminosities, $e$ and
$m_e$ are the electron charge and mass, $B_0=1$ G, $B$ is the magnetic
field in Gauss, $\nu_{0}=1.6 \times 10^{11}$ Hz  and $U_{0}=4.2 \times 10^{-13}$ erg cm$^{-3}$ are
 the CMB peak frequency and energy density at $z=0$, $\delta$ is the Doppler factor, and the
last part of each equation holds for equipartition conditions
\citep[$B\delta= 10^{-4}$ G;][]{jester05}.


To reproduce the UV - X-ray observations of knot A, 
a $B/\delta=5.5 \times 10^{-7}$ G is required ($\delta_{eq}=13.4$ assuming
equipartition). This determines, from the above equations and without any freedom, an IC/CMB component
peaking at at $\sim 10^{24.6}$ Hz with $\sim \nu f_\nu=10^{-11.7}$
erg/s/cm$^2$.  To demonstrate this, we plot in Figure \ref{knotA} a numerically
calculated SED taking into account electron energy  losses and the full Klein-Nishina  cross-section. 
Although this SED corresponds to equipartition conditions, numerical SEDs away from equipartition are practically identical for the required $B/\delta=5.5 \times 10^{-7}$ G.
 The level of the
IC/CMB emission at GeV energies violates the upper limit of the 3-10
GeV band at the $99.9\%$ level (Figure \ref{knotA}), {\sl ruling out the IC/CMB interpretation for the X-ray
  emission of knot A in 3C 273}. This is the main result of this work.


Abandoning the requirement that the UV - X-ray emission of
knot A is IC/CMB, we constrain $B/\delta > 1.3 \times 10^{-6}$ G, or $\delta_{eq}<9$
from the requirement that the IC/CMB emission
from knot A (broken line SED in Figure \ref{knotA}), does 
not overproduce the 3-10 GeV $95\%$  flux limit.

\subsection{Constraints from the A to D1 knot jet \label{AD1constraints}}

Radio polarization maps \citep{conway93} show that the jet magnetic
field runs roughly parallel to the jet from knot A all the way to knot
D1.  Beyond knot D1 the magnetic field turns abruptly to become
orthogonal to the jet axis, as one would expect from a shock that
decelerates the flow, and compresses the plasma.  The polarization is
suggestive of a jet that does not decelerate substantially from knot A
to knot D1, but decelerates efficiently past knot D1. It is thus
plausible that the flow from knot A to D1 is characterized by a single
Doppler factor, and that the magnetic field does not vary
significantly, as suggested by the fact that the equipartition
magnetic field of all knots is the same within a factor $<$2
\citep{jester05}.

\begin{figure}
  \includegraphics[width=3.4in]{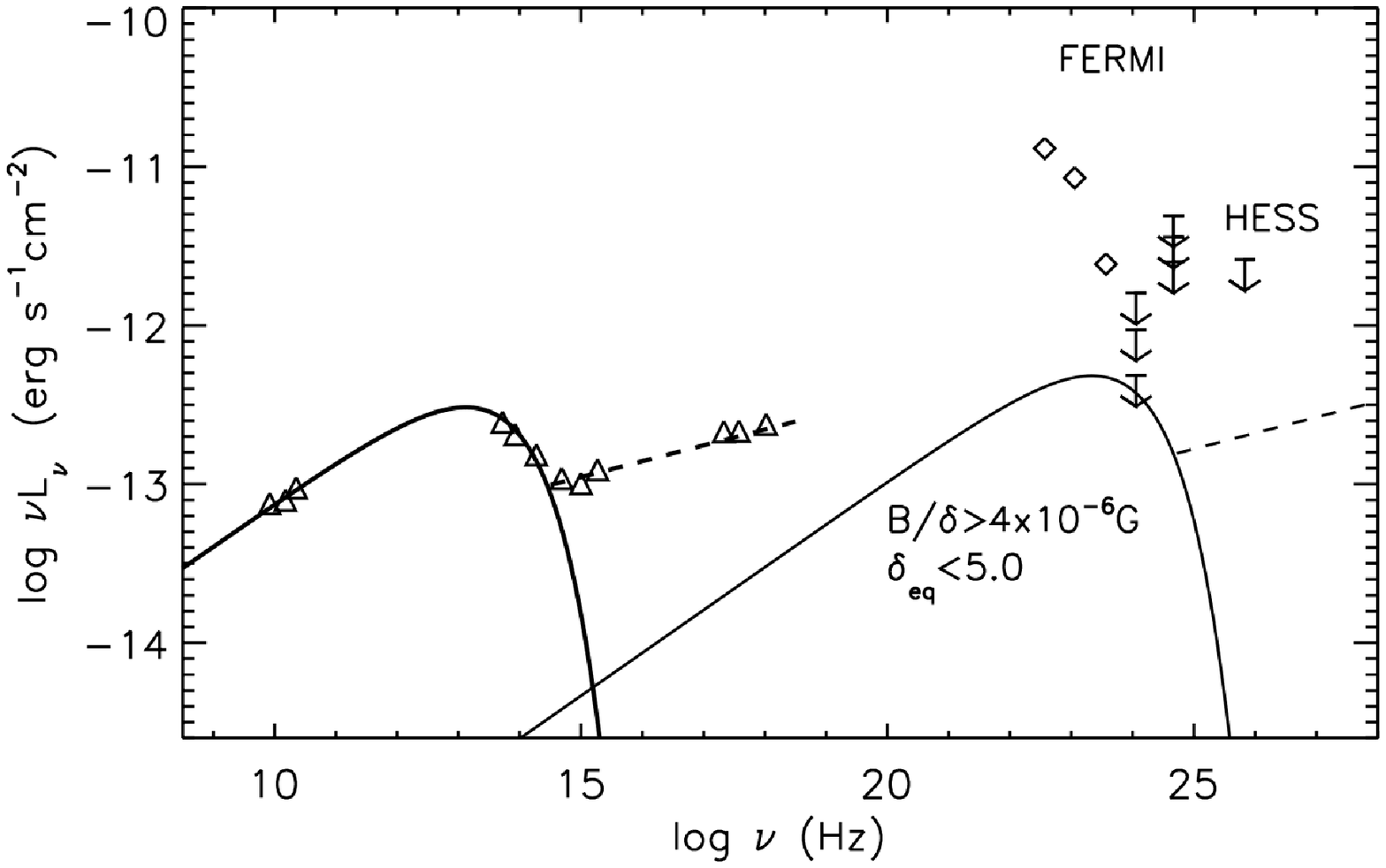}
  \caption{The SED of the jet from knot A to knot D1 (data from
    \citealt{uchiyama06} and \citealt{jester05,jester06}), including
    \emph{Fermi} measurements and upper limits described in \S
    \ref{sec:methods}, and a HESS upper limit \citep{aharonian05}.
    The thick solid line is a parametric fit of the synchrotron SED
    following \cite{uchiyama06} and GK06. The thick broken straight
    line is the SED of the UV - X-ray component, assumed to be of
    synchrotron nature.  The thin solid line, following the scalings
    of equations (\ref{eq:f},\ref{eq:L}) is the maximum amplitude the
    IC/CMB SED produced by the same electrons producing the
    synchrotron thick solid line SED can have without violating the
    3-10 GeV band \emph{Fermi} $95\%$ upper limit. The thin broken
    line is the IC/CMB SED that results from the same electrons that
    produce the UV-X-ray synchrotron emission.}
  \label{knotAtoD1}
\end{figure}

Based on the assumption that a single Doppler factor and magnetic
field describe the jet from knot A to D1 we can impose further
constraints. In Figure \ref{knotAtoD1} we plot the SED of the total
flux from knot A to D1, along with our \emph{Fermi} constraints. As
can be seen, to satisfy the 95\% 3-10 GeV band \emph{Fermi} limit we
require $B/\delta > 4.0 \times 10^{-6}$ G, or, assuming equipartition,
$\delta_{eq}<5$.  The existing shallow TeV limits \citep[3.9 h of HESS
  observations, no de-absorption applied;][]{aharonian05} do not
provide useful constraints, but future TeV observations with the
planned Cherenkov TeV Array (CTA) may be able to detect this
component.

\begin{figure}
  \includegraphics[width=3.2in]{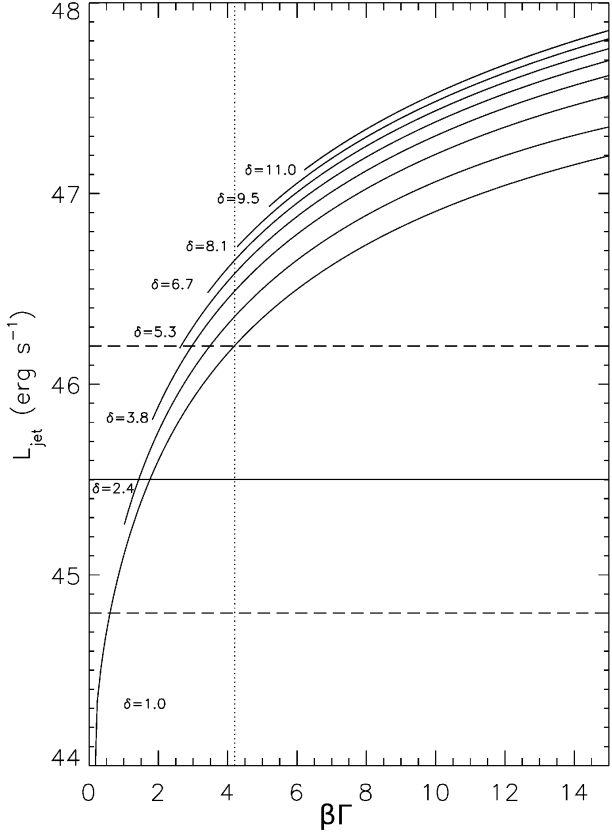}
  \caption{$L_{jet,min}$ as a function of $\beta\Gamma$
    for a range of $\delta$. The solid and broken horizontal lines
    represent the jet power estimate of $L_{jet}=10^{45.5\pm0.7}$ erg
    s$^{-1}$ from the X-ray cavity scaling. Jet configurations with
    $L_{jet,min} > 10^{46.2}$ erg s$^{-1}$ are disfavored, leading to an
    upper limit of $\Gamma\lesssim 4.2$ for the jet.}
  \label{power}
\end{figure}

\section{An upper limit on the bulk Lorentz factor}
\label{sec:bulkgamma}
We present here a model-dependent upper limit on $\Gamma$, based on
an estimate of the jet power $L_{jet}=10^{45.5\pm0.7}$ erg s$^{-1}$
\citep{mey11}
of 3C 273, derived from the scaling relation between kinetic
jet power estimated by the X-ray cavity method and the low frequency
radio lobe emission (\cite{cavagnolo10}, see also \cite{godfrey13} for a
different method extending the scaling to powerful jets). 
Assuming that the entire radio to X-ray emission of knot A comes from
the same region, the only frequency where an electron cooling break
can be manifested is either at $\nu_c=10^{13.5}$ Hz or $\nu_c\gtrsim10^{18}$ Hz, given
that no break is observed between the UV and X-ray observations
(assumed to come from a second synchrotron component).

 In the first case, the optical to X-ray emitting electrons are
 cooled, requiring an electron injection $q_{inj}(\gamma)\propto
 \gamma^{-1.5}$.  To calculate the minimum jet power $L_{jet,min}$
 corresponding to a given $\delta$ we calculate for a range of
 $\Gamma$ the magnetic field $B$ required for $\nu_c=10^{13.5}$ Hz.
 The Poynting power is then $L_B=\pi c R^2 \beta\Gamma^2 B^2/(8\pi)$,
 where $R$ is the radius of knot A. For this $B$ we then calculate the
 lepton power $L_{e}$required to produce the observed SED. With these,
 $L_{jet,min}=L_B+L_e$, because we do not include protons or thermal
 electrons.  In Figure \ref{power}, we show curves of $L_{jet,min}$ as
 a function of $\beta\Gamma$ for a range of $\delta$ Configurations
 that require $L_{jet,min} > 10^{46.2}$ erg/s are disfavored, leading
 to $\Gamma\approx \beta \Gamma<4.2$

 In the second case the optical to X-ray emitting electrons escape the
 emission region before cooling and electron injection is steeper,
 $n_{inj}(\gamma)\propto \gamma^{-2.5}$.  This second case of no
 cooling up to $10^{18}$ Hz requires that a region significantly
 smaller than the optical jet lateral size ($\sim 1$ kpc) is
 responsible for the UV to the X-ray emission: for example, for
 $\delta=\Gamma=5$ the maximum size of this emitting region is $\sim
 100$ pc, corresponding to a variability timescale of $\sim 70$ years.

\section{Discussion}
\label{sec:discussion}

Using upper limits to the \emph{Fermi} flux of the LSJ of 3C 273, we
rule out IC/CMB being the X-ray emission mechanism of knot A, the
X-ray dominant knot of the LSJ.  This result does not depend on any
assumptions of equipartition or jet content and is, therefore,
robust. Assuming equipartition and a steady flow from knot A to knot
D1, as suggested by observations, we set an upper limit to the jet
Doppler factor, $\delta\leq 5$. Finally, adopting an upper limit to
the jet power derived from the X-ray cavity scaling, we find
$\Gamma\lesssim 4.2$. We note that the IC/CMB mechanism for the rest
of the knots, all downstream of knot A, has been discounted on
spectral (X-ray spectrum significantly steeper than the radio) or
morphological (radio and optical emission not co-located) grounds
\citep{jester05,jester06,jester07}.

Our result leaves as the only alternative a synchrotron nature for the
X-ray emission.  This means that \emph{in situ} particle acceleration
takes place that accelerates electrons at least up to $\sim 30-100$
TeV. It is not clear what particle acceleration mechanism produces
this second EED.  If we assume that this population cools before it
escapes the emission region, a very hard electron injection is
required ($n_{inj}(\gamma)\propto \gamma^{-1.5}$).  On the other hand,
if the electrons escape the emission region uncooled, a steeper
electron injection is needed ($n_{inj}(\gamma)\propto \gamma^{-2.5}$),
but this requires that the emission region is significantly smaller
than 1 kpc. Assuming $\delta=\Gamma=5$ this corresponds to 100 pc and
to a variability timescale of $\sim 70$ years \citep[note that X-ray
  variability with a timescale of a few years has been observed for a
  kpc scale knot in the LSJ of Pictor A;][]{marshall10}.  We finally
note that while IC/CMB appears to be ruled out in 3C 273, it is
possible that other powerful LSJs produce X-rays through IC/CMB.


\begin{acknowledgments}
EM acknowledges support from \emph{Fermi} grant NNX10AO42G. MG acknowledges
support from  \emph{Fermi} grant NNX12AF01G. 
\end{acknowledgments}


\end{document}